\documentclass[12pt]{article}
\usepackage{a4wide}

\newcommand{\de}{\delta}

\newcommand{\cL}{{\cal L}}
\newcommand{\M}{{\cal M}}
\newcommand{\N}{{\bf N}}
\newcommand{\R}{{\bf R}}

\begin{document}

\title{\vskip -70pt
  \begin{flushright}
    {\normalsize{DAMTP 97-14.\\
    hep-th/9703024.}} 
  \end{flushright}
  \vskip 15pt {\bf \Large Construction of supercharges for the
one-dimensional supersymmetric nonlinear sigma model}} 

\author{ A. J. Macfarlane \\
{\normalsize E-mail: {\tt A.J.Macfarlane@damtp.cam.ac.uk}} \\[10pt]
and A. J. Mountain\thanks{Present address :
The Blackett Laboratory, Imperial College, Prince
Consort Road, London SW7 2BZ, United Kingdom} \\
{\normalsize E-mail: {\tt A.Mountain@ic.ac.uk}} \\ [10pt]
{\normalsize \sl Department of Applied Mathematics and Theoretical
Physics,} \\  
{\normalsize \sl University of Cambridge, Silver St., Cambridge CB3 9EW,
United Kingdom.}\\[10pt]}

\date{May 1999}

\maketitle

\vskip60pt

\begin{abstract}
This paper addresses an issue essential to the study of hidden
supersymmetries (meaning here ones that do not close on the
Hamiltonian) for one-dimensional non-linear supersymmetric sigma
models. The issue relates to ambiguities, due to partial integrations in
superspace, both in the actual definition of these  supersymmetries and
in the Noether definition of the associated supercharges. The unique
consistent forms of both these definitions have to be determined
simultaneously by a process that adjusts the former definitions so that
the associated supercharges do indeed correctly generate them with the
aid of the canonical formalism. The paper explains and illustrates these
matters and gives some new results.
\end{abstract}

\newpage
\section{Introduction}
 
This paper is concerned with the study of certain aspects of
one-dimensional supersymmetric sigma models that have not received
systematic study elsewhere.  It has been well-understood for a long
time how superfield methods for such models provide an efficient and clear
description of their natural, $N=1$, supersymmetry.  This is generated
by a conserved Hermitian supercharge $Q$, which is constructed
directly in a problem-free fashion using Noether's theorem and which
``closes'' on the Hamiltonian of the theory via $H=Q^2$.  Much the
same applies to $N$-extended supersymmetries when extended superfields
are used, the Hermitian supercharges $Q_a, \; a=1, 2 \dots N$, closing on
the Hamiltonian in this context via $\{ Q_a, Q_b \} = 2\delta_{ab}
H$.  In this paper, we study the conditions under which further
`hidden' supersymmetries are present in such models and the consistent
determination of the supercharges which generate them.

The main aspects of our work for which novelty and importance is
claimed stem in large part from our relaxation of the requirement that
all supersymmetries should close (as noted for $N=1$ and extended
supersymmetries) upon the Hamiltonian.  Rather, when a hidden
supersymmetry with generator $Q'$ has been found in a consistent
fashion,  it is then necessary to compute the constant of the motion
$K$ that follows via $K=Q'^{2}$,{\it i.e.} the algebra of hidden
supersymmetries has to be determined after the supersymmetries themselves
have been discovered.  For an example of this
situation, one that is so simple as to be free of technical
difficulties, see \cite{deJongheMPandvanH}. It is the present widening of the
context which gives rise to ambiguities in the path from the action
$S$ of the theory to hidden supersymmetries and their generators,
causing problems that are of a troublesome sort and of widespread
occurrence.  We turn next to explaining the nature of the ambiguities
and to describing the procedure necessary and sufficient for their
unique resolution.

First we consider the contrasting situations for natural and hidden
supersymmetries.  For the former the transformations in question are
known a priori and yield an exactly calculable total divergence for
$\delta S$, so that Noether's theorem produces (what always turns out to be)
the correct supercharge $Q$ without ambiguity.  In the latter case one
is trying to {\it determine} hidden supersymmetries, that is to determine
conditions on the unknown tensors that feature in some ansatz for them,
in such a way that $\delta S=0$ follows.  Some partial integrations in
superspace are essential in this process, but wide variations in these,
all of which do yield $\delta S=0$, are possible.  They lead to
different conditions on the unknown tensors, different total divergence
expressions for $\delta S$, and hence (via Noether's
theorem) to different expressions for the corresponding
supercharges. How does one chose the correct procedure and identify the
correct results? Clearly a criterion from {\it outside} the calculation
just described is called for and, of course, one exists.

The criterion for the correct resolution of these ambiguities is easily
stated. One must perform the integrations in superspace which determine
the explicit  form of the hidden supersymmetry transformations giving
invariances of the action in such a way that the corresponding Noether
supercharge generates via the canonical formalism exactly the same
supersymmetry transformations. For a natural supersymmetry, where the
the transformations are known from the outset, this happens routinely
without problem, as already noted. Although it was easy to state our
criterion, there is, in the contexts of interest in the present paper,
no systematic way to achieve its implementation. Rather, one has to
proceed starting from an initial form of the hidden supersymmetry that
does give $\delta S=0$, computing the corresponding $Q$, observing if
(and of course exactly how) it fails to give back the original
supersymmetry transformations and adjusting these transformations until
one reaches the goal required by the consistency criterion. We emphasise
that there is no other way to escape the problems described except in
very simple situations. The illustration presented in Section
\ref{examplesect} gives a good impression of the nature and treatment of
the problem in a modestly difficult case. It should be apparent even
here that that we are raising an important issue. 

We continue our introduction with some background material on sigma
models and supersymmetries including extended and hidden ones, in part
amplifying the summary just given of the problems on which the paper
focuses. A sigma model is an action for dynamical fields considered as
maps from a spacetime to a target manifold. In the case where the
spacetime is $(1+1)$-dimensional, the sigma-model can be seen to provide
the action for a string world-sheet and to describe the propagation of a
string in the target manifold. There is a generalisation to a
supersymmetric worldsheet with corresponding target space supersymmetry
which describes the propagation of a superstring. Indeed, for these
reasons there has been extensive study of such low-dimensional
sigma-models. Many interesting features of sigma-models can be
illustrated by considering the case in which the spacetime is
one-dimensional and parameterised by a real time co-ordinate. This leads
to the action for a particle propagating on the target manifold, and the
supersymmetric generalisation is clear. It is with such models that we
are concerned here. 

We shall study the one-dimensional nonlinear sigma-model with
$N=1$ supersymmetry in the case in which the target manifold is a
principal fibre bundle $P(\M,G)$. We are particularly interested in the
case where $G$ is a compact Lie group. By considering a sigma model involving
bosonic $N=1$ superfields valued in $\M$ and fermionic superfields
valued in $G$, we arrive at an action for a supersymmetric point
particle with internal ``colour'' spin degrees of freedom transforming
under $G$. To complete
the picture, we introduce a background Yang-Mills field as the curvature
of a connection on $P$. The colour degrees of freedom are minimally
coupled to the bosonic superfields via the gauge potential.

Such a model, by virtue of its superspace construction exhibits an
explicit $N=1$ supersymmetry. There exists a (fermionic)
supercharge $Q_0$ which, upon use of the canonical formalism, generates
the canonical $N=1$ supersymmetry transformations. The explicit nature
of this supersymmetry allows the definition of
$Q_0$ via the Noether procedure to proceed without the ambiguities
mentioned above. Further use of the classical canonical formalism allows
us to calculate the ``square'' of $Q_0$ via the Poisson-Dirac bracket, giving
the Hamiltonian for the theory as $\{Q_0\, ,\,Q_0\}=-2iH$ (in the
quantum theory there is an extra factor of $i$ on the right hand
side). We say that $Q_0$ ``closes'' on the Hamiltonian.

\subsection{Extended supersymmetry}
\label{extendedsusysect}

There is by now a very
large body of work including \cite{ColesandP, HoweandT, Rietdijk,
  MacfarlaneSpin, LindenMandvanH, Papadopoulos, EvansandM} on $N = 1$
supersymmetric quantum-mechanical models in which there is a single
Hermitian supercharge that closes on the
Hamiltonian. Clearly, great importance is attached to the search for
additional supersymmetries in such models. An additional supersymmetry
is a set of transformations which leave the action invariant and commute
with the original supersymmetry transformations. These will in turn generate an
extra (fermionic) supercharge $Q'$, say, which, upon use of the
Poisson-Dirac bracket, will then satisfy $\{Q_0 \, , \, Q'\} = 0$. The
bosonic quantity constructed from $Q'$ by $K = \frac{i}{2} \{Q' \, , \,
Q'\}$ can be seen, upon use of the Jacobi identity, to be time-invariant. 

There is a large body of literature on so-called extended
supersymmetry. This is defined as a set of extra supersymmetries of the
type described above, all with $K=H$, leading to the algebra
\begin{equation}
  \{Q_\alpha , Q_\beta \} = 2 \delta_{\alpha \beta} H \quad,
\end{equation}
in quantum mechanics, or classically, in terms of Poisson brackets,
\begin{equation}
  \{Q_\alpha , Q_\beta \} = -2i \delta_{\alpha \beta} H \quad.
\end{equation}
There are two possibilities for generating such an extra supersymmetry
of the one-\linebreak dimensional supersymmetric nonlinear
sigma-model. The first is related to an endomorphism symmetry of the
tangent bundle generated by a complex structure. A complex structure $I$
is a $(1,1)$-tensor on $\M$ which gives a closed two-form on $\M$ by
$\omega = gI$, where $g$ is the metric. To be a complex structure, a
tensor $I$ must be covariantly constant and satisfy $I^2=-{\bf1}$,
with ${\bf 1}$ the identity. A manifold with such a structure is said to
be K\"ahler and must be of real dimension $2n$ for $n \in \N$. This
allows the supersymmetry to be extended from $N=1$ to $N=2$. In the case
where there are three complex structures $I$, $J$ and $K$ which satisfy
the quaternion algebra, the manifold is hyperK\"ahler and must have real
dimension $4n$ for $n \in \N$. The Poisson-Dirac brackets among the set
of corresponding supercharges vanish, leaving an $N=4$ extended
supersymmetry. 

The second type of extended supersymmetry is generated from
supersymmetries bet\-ween the fields on the base manifold $\M$ and
fields on the fibre $G$. This requires $\M$ and $G$ to have the same
dimension. Maps between the two can be interpreted as vielbeine. The
complex structures (see above) relate the dynamical bosons on $\M$ to
the fermions on $\M$ via supersymmetry. The maps between $\M$ and $G$
allow us to relate, via supersymmetry, the dynamical bosons on $\M$ to
the fermions on $G$. Thus these maps play a similar role to that of
the complex structures and satisfy similar conditions. 

A derivation is given by Coles and Papadopoulos \cite{ColesandP} of
a complete set of conditions sufficient for invariance of the action
under both types of supersymmetry for the one-dimensional supersymmetric
nonlinear sigma model.

\subsection{Hidden supersymmetries}

In this paper, we wish to study the generalisation of
extended supersymmetry found by relaxing the condition that the extra
supercharges close on the Hamiltonian. Use of the Jacobi identity
allows us to see that any quantity obtained as a square of such a
generalised supercharge commutes with the Hamiltonian and is itself a (bosonic)
constant of the motion. We describe the conditions under which such
generalised ``hidden'' supersymmetries exist in one-dimensional
supersymmetric sigma-models. 

Supersymmetry algebras which, as in \cite{deJongheMPandvanH,
DHokerandV}, do not close on $H$, rather on different
important constants of the motion, have been displayed for several
theories, including the motion of a spinning particle in a Dirac monopole
\cite{deJongheMPandvanH}, Kerr-Newman \cite{GibbonsRandvanH} and
Taub-NUT \cite{vanHolten, Visinescu} background. These involve at
least one, one and four additional supercharges respectively. A similar
treatment has been applied to the case of a particle with both spin and
colour degrees of freedom in a background Yang-Mills field in
\cite{AJMandAJM}. The examples cited all involve the use, in an
essential way, of Killing-Yano tensors, a topic comprehensively
discussed by Tanimoto \cite{Tanimoto} for a general curved plus
electromagnetic background. These are essentially generalisations of the
complex structures which are used to generate extended supersymmetry
(see above). Due to the fact that the associated supercharges are not
required to close on the Hamiltonian, they satisfy less restrictive
conditions and a much wider class of manifolds admit such
supersymmetries. There are also suitable generalisations of the second
class of extended supersymmetries described above, as illustrated in
\cite{AJMandAJM}. These involve supersymmetries between the dynamical
bosons on $\M$ and the dynamical fermions on $G$ and are generalisations
of those described in Section 
\ref{extendedsusysect}.

\subsection{Construction of supercharges}

In an important paper on the supersymmetries of the one-dimensional
supersymmetric nonlinear sigma model, Coles and Papadopoulos
\cite{ColesandP} provide a list of conditions sufficient for the
invariance, $\delta S = 0$, of the action $S$. We say ``sufficient''
here because, as is noted in \cite{ColesandP}, the list contains
ambiguities due to the role played by partial integrations in
superspace. In relation to invariance, this may well not be of major
significance. However, for our purpose, detailed analysis and treatment of such
ambiguities is of paramount importance. We require the explicit
construction of the supercharge $Q$ associated with each set of
supersymmetry transformations $\cal J$ under which $\delta S = 0$, and
we use Noether's theorem to perform it. The ambiguities mentioned above
assume real significance at this point since we may well (and in general
do) have $\delta S = 0$ and $\delta L \neq 0$, since a total time
derivative in $\delta L$ does not contribute to $\delta S$. As the
construction of the associated supercharges involves $\delta L$, related
ambiguities affect the supercharges and manifest themselves when a
plausible expression for $Q$ is used within the canonical formalism of
the theory to calculate the supersymmetry transformations that $Q$
generates. Consistency requires that the results of these calculations
should coincide with the original transformations $\cal J$ but they may
well fail to do so. In all but very simple contexts, it is in general a
highly non-trivial matter to handle the partial integrations in exactly
the fashion that is required to achieve consistency.

The paper is organised as follows. In Section \ref{setupsect}, we
describe the supersymmetric formalism of the one-dimensional
supersymmetric sigma-model. We describe the canonical quantisation of
such a model and the $N=1$ supersymmetry algebra in Section
\ref{canonsect}. Then, in Section \ref{Qsect}, we discuss the existence
of additional supersymmetries of the model and construct Noether charges
for the
original and additional supersymmetries. Having exposed the ambiguities
involved in calculation of the supercharges,
we determine the conditions for invariance of $S$ in the form required
for the consistent construction of the supercharges and perform that
construction explicitly. This culminates in the display of the
conditions for invariance of the action that embody consistent
treatment of supercharges, for which explicit expressions are given.
Canonical equations in alternative form, useful for some purposes, are
mentioned in Section \ref{simpsect}. In Section \ref{examplesect}, we
describe an example that
makes fully explicit the nature and resolution of the problem of
consistent calculation of $Q$.

For simplicity, we present our discussion in the language of classical
mechanics; the extension of it to the quantum case proceeds
in straightforward fashion.

\section{The one-dimensional supersymmetric sigma-model}
\label{setupsect}

Sigma-models are theories involving fields considered as maps from
spacetime $S$ to a target manifold $\M$. The case $S=\R^{(1,1)}$ has
been much-studied as this describes the propagation of a string in the
background manifold $\M$. It is also interesting to consider the case
of $S=\R$, parameterised by time. This gives a theory of quantum
mechanics on the manifold $\M$. If $\M$ is $n$-dimensional, there
exist co-ordinates on $\M$ such that a given field $\Phi$, say, is
composed of $n$ maps
\begin{equation}
  \Phi^i : S \rightarrow \M \quad.
\end{equation}
This will be taken to be an $N=1$ superfield with bosonic components
$x^i$ and fermionic components $\psi^i$. We introduce co-ordinates
$(t,\theta)$ on the one-dimensional, $N=1$ superspace, with $t$ real and
$\theta$ a Grassmann parameter and write
\begin{equation}
  \Phi^i = x^i(t) + i \theta \psi^i(t) \quad.
  \label{bosonfield}
\end{equation}
The model studied in this paper is that where the target space is a
principal bundle $P(\M,G)$.  Dynamical fermions, valued in the fibre,
are defined by the fermionic superfield $\Lambda^\alpha =
\lambda^\alpha + \theta F^\alpha$, where $\lambda^\alpha$ are
fermionic and the $F^\alpha$ are auxiliary bosonic fields. We use the
superderivative $D=\partial_\theta - i\theta \partial_t$ to write down
an action for the fields $\Phi^i$ and $\Lambda^\alpha$, with minimal
coupling:
\begin{eqnarray}
  S  &=&  \int \,d \theta \,dt \cL \quad,
  \label{action} \\
  \cL & = & \frac{1}{2} \left( i g_{ij}(\Phi) \dot{\Phi^i} D \Phi^j -
    \frac{1}{3} c_{ijk}(\Phi) D \Phi^i D \Phi^j D \Phi^k + h_{\alpha
    \beta} \Lambda^\alpha \nabla \Lambda^\beta \right) \quad.
  \label{superL}
\end{eqnarray}
This involves the covariant derivative $\nabla \Lambda^\alpha = D
\Lambda^\alpha + {{A_i}^\alpha}_\beta(\Phi) D \Phi^i \Lambda^\beta$,
where ${{A_i}^\alpha}_\beta$ is a gauge connection with field strength
\begin{equation}
  {{F_{ij}}^\alpha}_\beta = \partial_i {{A_j}^\alpha}_{\beta} -
    \partial_j {{A_i}^\alpha}_{\beta} + {{A_i}^\alpha}_\gamma
    {{A_j}^\gamma}_\beta - {{A_j}^\alpha}_\gamma {{A_i}^\gamma}_\beta
    \quad. 
  \label{fieldstrength}
\end{equation}
The second term of (\ref{superL}) arises from a partial integration in
superspace of the Wess-Zumino term $\frac{i}{2} b_{ij} \dot{\Phi^i} D
\Phi^j$, so that the $c_{ijk}$ are the components of the 3-form $c =
-\frac{3}{2} d b$. $c$ can be interpreted as the torsion of the manifold
$\M$. The fibre is a compact Lie group and hence the metric on the fibre
can be taken to be $h_{\alpha \beta} = \de_{\alpha \beta}$.  Writing
$\cL = K + \theta L$, we have
\begin{eqnarray}
K & = & \frac{1}{2} \left( - g_{ij} \dot{x}^i \psi^j + \frac{i}{3} c_{ijk} \psi^i
  \psi^j \psi^k + h_{\alpha \beta} \lambda^\alpha F^\beta - i h_{\alpha \beta}
  \lambda^\alpha {{A_i}^\beta}_\gamma \lambda^{\gamma} \psi^i \right) \quad,
  \label{Kdef} \\
L & = & \frac{1}{2} \left( g_{ij} \dot{x}^i {\dot{x}}^j + i g_{ij} \psi^i
  {\dot{\psi}}^j - i g_{ij,k} \psi^k {\dot{x}}^i \psi^j - i c_{ijk} \dot{x}^i
  \psi^j \psi^k \right. \nonumber \\
  & & \mbox{} - \frac{1}{3} c_{ijk,n} \psi^n \psi^i \psi^j \psi^k + i
  h_{\alpha \beta} \lambda^\alpha {\dot{\lambda}}^\beta + h_{\alpha
  \beta} F^\alpha F^\beta + i h_{\alpha \beta,i} \psi^i \lambda^\alpha
  F^\beta \nonumber \\
  & & \mbox{} - i h_{\alpha \beta} {{A_i}^\beta}_\gamma \left( F^\alpha
  \lambda^\gamma - \lambda^\alpha F^\gamma \right) \psi^i + h_{\alpha \beta}
  {{A_i}^\beta}_{\gamma,k} \psi^k \lambda^\alpha \lambda^\gamma \psi^i
  \nonumber \\
  & & \mbox{} \left. + i h_{\alpha \beta} {{A_i}^\beta}_\gamma
  \lambda^\alpha \lambda^\gamma {\dot{x}}^i + h_{\alpha \beta,j}
  \psi^j {{A_i}^\beta}_\gamma \lambda^\alpha \lambda^\gamma \psi^i
  \right) \quad.
  \label{Ldef}
\end{eqnarray}
The fields $F^\alpha$ are non-dynamical and can be eliminated using
their Euler-Lagrange equations. Using this, the condition $\nabla h =
0$ and the fact that $c$ is a closed 3-form, $d c =0$, we have
\begin{eqnarray}
L & = & \frac{1}{2} \left( g_{ij} \dot{x}^i \dot{x}^j + i g_{ij} \psi^i
    {\dot{\psi}}^j + i g_{ij,k} \psi^j \psi^k \dot{x}^i - i c_{ijk}
    \dot{x}^i \psi^j \psi^k + i h_{\alpha \beta} \lambda^\alpha
    {\dot{\lambda}}^\beta \right) \nonumber \\
  & & \mbox{} + \frac{i}{2} A_{i \alpha \beta} \lambda^\alpha
    \lambda^\beta {\dot{x}}^i + \frac{1}{4} F_{ij \alpha \beta}
    \lambda^\alpha \lambda^\beta \psi^i \psi^j \quad.
  \label{Ldef2}
\end{eqnarray}
The Lagrangian (\ref{superL}) is written in terms of $N=1$ superfields
and, as such, has an explicit $N=1$ supersymmetry given in terms of a
Grassmann parameter $\epsilon$ by
\begin{eqnarray}
  \de t = -i \epsilon \theta \quad, \qquad \de \theta = - \epsilon \quad.
  \label{Q0transfns}
\end{eqnarray}
The generator of these transformations is the supercharge $Q_0$. This is
fermionic and has an explicit expression in terms of the dynamical
fields of the model. We can give the canonical formalism for the
Lagrangian in the usual way via the Poisson-Dirac bracket and show that
the supercharge generates the correct transformations of the
dynamical variables via the Poisson-Dirac bracket. This will be
described in detail below. The square of the supercharge, computed via
the Poisson-Dirac bracket, gives the classical Hamiltonian by
\begin{equation}
  \{Q_0 \, , \, Q_0\} = -2iH \quad.
\end{equation}
The analogue of this statement in quantum mechanics is with the
right-hand side multiplied by $i$. It should be noted that this
procedure does generate the correct Hamiltonian; this can be seen by
calculating it in the conventional way as
\begin{equation}
  H = \sum \dot{X} \frac{\partial L}{\partial \dot{X}} - L \quad,
\end{equation}
where the sum is over all dynamical variables. The time evolution of
an arbitrary quantity $K$ is then given by
\begin{equation}
  \frac{dK}{dt} = \{K\, , \,H\} \quad.
\end{equation}
This leads to the important observation that any quantity $Q'$ which
satisfies $\{Q_0\, , \, Q'\}=0$ generates a constant of the motion via
$K=-\frac{i}{2} \{Q'\, , \,Q'\}$ because $\{K\, , \,H\}=0$ upon use of
the Jacobi identity.

\section{Canonical Quantisation}
\label{canonsect}

The quantisation of this model follows familiar lines. From
(\ref{Ldef2}) we derive the following conjugate momenta
\begin{eqnarray}
  p_i = \frac{\partial L}{\partial {\dot{x}}^i} & = & g_{ij}{\dot{x}}^j
    + \frac{i}{2}g_{ij,k} \psi^j \psi^k + \frac{i}{2} A_{i \alpha \beta}
    \lambda^\alpha \lambda^\beta - \frac{i}{2} c_{ijk} \psi^j \psi^k
    \quad, \\ 
  \tau_i = \frac{\partial L}{\partial {\dot{\psi}}^i} & = &
    - \frac{i}{2}g_{ij} \psi^j \quad, \\
  \xi_\alpha = \frac{\partial L}{\partial {\dot{\lambda}}^\alpha} & = &
    - \frac{i}{2} h_{\alpha \beta} \lambda^\beta \quad.
  \label{conjmom}
\end{eqnarray}
Thus we have two constraint functions 
\begin{eqnarray}
  \eta_i & = & \tau_i + \frac{i}{2} g_{ij} \psi^j \quad, \\
  \sigma_\alpha & = & \xi_\alpha + \frac{i}{2} h_{\alpha \beta}
    \lambda^\beta \quad.
  \label{constraintfns}
\end{eqnarray}
We use the fundamental brackets
\begin{equation}
  \begin{array}{lll}
  {\{x^i , p_j\} = {\delta^i}_j}\quad, & {\{\psi^i , \tau_j\} =
    -{\delta^i}_j}\quad, & {\{\lambda^\alpha , \xi_\beta\} =
    -{\delta^\alpha}_\beta} \quad,
  \end{array} 
\end{equation}
to obtain
\begin{eqnarray}
  \{\eta_i , \eta_j\} & = & - i g_{ij} \quad, \\
  \{\sigma_\alpha , \sigma_\beta\} & = & - i h_{\alpha \beta} \quad,
\end{eqnarray}
and define the Dirac bracket $\{A,B\}^*$ by
\begin{equation}
  \{A,B\}^* = \{A,B\} - \{A,\eta_i\} i g^{ij} \{\eta_j,B\} -
    \{A,\sigma_\alpha\} i h^{\alpha \beta} \{\sigma_\beta,B\} \quad.
  \label{Dbracket}
\end{equation}
Since, from now on, all brackets will be Dirac brackets, the asterisk
is left implicit. We also work with the covariant momentum
\begin{equation}
\pi_i = g_{ij} {\dot{x}}^j = p_i - \frac{i}{2}g_{ij,k} \psi^j \psi^k -
  \frac{i}{2} A_{i \alpha \beta} \lambda^\alpha \lambda^\beta +
  \frac{i}{2} c_{ijk} \psi^j \psi^k \quad.
  \label{covmom}
\end{equation}
We then have the following canonical equations
\begin{equation}
  \begin{array}{lll}
  \{x^i,x^j\} = 0 & \{x^i,\pi_j\} = \delta^i_j & \{x^i,\psi^j\} = 0
    \quad, \\ \vspace{3pt}
  \{x^i,\lambda^\alpha\} = 0 & \{\psi^i,\pi_j\} = - {\Gamma^i}_{jk}
    \psi^k -{c^i}_{jk} \psi^k & \{\lambda^\alpha,\pi_i\} = -
    {{A_i}^\alpha}_\beta \lambda^\beta \quad, \\ \vspace{3pt}
  \{\psi^i,\psi^j\} = -i g^{ij} & \{\lambda^\alpha,\lambda^\beta\} =
    -i h^{\alpha \beta} & \{\psi^i,\lambda^\alpha\} = 0 \quad, \\ 
  \multicolumn{3}{c}{\{\pi_i,\pi_j\} = \frac{i}{2} R_{ijpq} \psi^p
    \psi^q + \frac{i}{2} F_{ij \alpha \beta} \lambda^\alpha
    \lambda^\beta + i \nabla_{[j} c_{i]pq} \psi^p \psi^q - i c_{inp}
    {{c_j}^n}_q \psi^p \psi^q} \quad,
  \end{array}
  \label{1stbrackets}
\end{equation}
where $F_{ij \alpha \beta}$ is defined in (\ref{fieldstrength}) and
\begin{equation}
  R_{ijpq} = g_{in} \left( \partial_p {\Gamma^n}_{jq} - \partial_q
    {\Gamma^n}_{jp} + {\Gamma^n}_{kp} {\Gamma^k}_{jq} -
    {\Gamma^n}_{kq} {\Gamma^k}_{jp} \right) \quad.
\end{equation}
We define a generalised connection ${\tilde{\Gamma}^i}_{jk} =
{\Gamma^i}_{jk} + {c^i}_{jk}$ and a corresponding generalised
curvature $\tilde{R}_{ijpq}$, defined from $\tilde{\Gamma}$ as $R$ is
from $\Gamma$, to obtain
\begin{equation}
  \tilde{R}_{ijpq} = R_{ijpq} + 2 \nabla_{[j} c_{i]pq} - 2 c_{inp}
    {{c_j}^n}_q \quad,
\end{equation}
so that the brackets (\ref{1stbrackets}) lead to the form for the
general Dirac bracket
\begin{eqnarray}
  \{A,B\} & = & \frac{\partial A}{\partial x^i} \frac{\partial
    B}{\partial \pi_i} -\frac{\partial A}{\partial \pi_i}
    \frac{\partial B}{\partial x^i} \nonumber \\
  & + & \frac{\partial A}{\partial \pi_i} \frac{\partial B}{\partial
    \pi_j} \left( \frac{i}{2} \tilde{R}_{ijpq} \psi^p \psi^q +
    \frac{i}{2} F_{ij \alpha \beta} \lambda^\alpha \lambda^\beta
    \right) \nonumber \\
  & - & \left({(-)}^b \frac{\partial A}{\partial \pi_i}
    \frac{\partial B}{\partial \psi^j} - {(-)}^{a+b} \frac{\partial
    A}{\partial \psi^j} \frac{\partial B}{\partial \pi_i} \right)
    {\tilde{\Gamma}^j}_{ik} \psi^k \nonumber \\
  & - & \left( {(-)}^b \frac{\partial A}{\partial \pi_i} \frac{\partial
    B}{\partial \lambda^\alpha} - {(-)}^{a+b} \frac{\partial
    A}{\partial \lambda^\alpha} \frac{\partial B}{\partial \pi_i}
    \right) {{A_i}^\alpha}_\beta \lambda^\beta \nonumber \\
  & + & i {(-)}^a \frac{\partial A}{\partial \psi^i}
    \frac{\partial B}{\partial \psi^j} g^{ij} + i {(-)}^a
    \frac{\partial A}{\partial \lambda^\alpha}
    \frac{\partial B}{\partial \lambda^\beta} h^{\alpha \beta} \quad,
  \label{bracket}
\end{eqnarray}
where $a$ and $b$ are the Grassmann parities of $A$ and $B$
respectively.

\section{Construction of Supercharges}
\label{Qsect}

\subsection{The construction of Noether charges}
\label{constructsect}

We construct supercharges from general superfield transformations
$\delta \Phi^i$ and $\delta \Lambda^\alpha$ which leave the superfield
action (\ref{action}) invariant. By Noether's theorem, there must exist
a conserved supercharge which generates each such set of
transformations. Below is a description of
the construction of these supercharges. An explicit example is given
in Section \ref{examplesect}, which is intended to illustrate the
principles involved. We set out from the action (\ref{action})
\begin{eqnarray}
  S & = & \int \,d \theta \,dt \cL = \int \,dt L \quad, \nonumber \\
  \cL & \equiv & \cL (\Phi, D \Phi, \dot{\Phi}, \Lambda, D
    \Lambda) \quad. \nonumber 
\end{eqnarray}
We now calculate $\delta L$ in two ways. First, we apply the
transformations of $\Phi$ and $\Lambda$ directly to the action and
integrate over the Grassmann variable $\theta$ to get $\delta L$ in
the form
\begin{equation}
  \delta L = \frac{dJ}{dt} + \mbox{\{other terms\}} \quad.
  \label{deltaL1}
\end{equation}
The ``other terms'' are required to vanish, leaving just the total
derivative. This requirement imposes a set of conditions on the
initial superfield transformations, which we examine in Section
\ref{extrasusysect}. This leaves $\delta L$ as a total derivative
which ensures that the action is invariant. We can also get an
expression for $\delta L$ within the standard procedure for Noether's
theorem in the form
\begin{equation}
  \delta L = \partial_t \left( \sum_X \delta X \frac{\partial
    L}{\partial \dot{X}} \right) \quad,
  \label{deltaL2}
\end{equation}
where the sum is over the dynamical variables $x$, $\psi$ and
$\lambda$. Equating (\ref{deltaL1}) and (\ref{deltaL2}), we have a
time-invariant supercharge $Q$ given by
\begin{eqnarray}
  i \epsilon Q & = & \sum_X \delta X \frac{\partial L}{\partial \dot{X}} -
    J \quad, \label{schargedef} \\ 
  \partial_t Q & = & 0 \quad.
\end{eqnarray}
In the above expression for $Q$, $\epsilon$ is the constant,
Grassmann-odd parameter which appears in the supersymmetry
transformations. As indicated in the Introduction, difficulties in
implementing the procedure outlined arise in supersymmetric theories
from the possibility of integrating by parts in superspace, so that the
separation of (\ref{deltaL1}) into the two indicated pieces is not
unique. Given a version of (\ref{deltaL1}), a further such integration
may change both $J$ and the ``other terms'' and hence the conditions
under which the latter vanish. 

\subsection{The fundamental supercharge and the Hamiltonian}

The Lagrangian (\ref{superL}) is of the form $\cL \equiv \cL (\Phi ,
D \Phi , \dot{\Phi} , \Lambda , D \Lambda)$.  Consequently, it is
manifestly invariant under the original supersymmetry transformations
of the theory (\ref{Q0transfns}), which are realised on the
superfields as
\begin{eqnarray}
  \delta \Phi^i & = & -\epsilon {\cal D} \Phi^i \quad, \\
  \delta \Lambda^\alpha & = & -\epsilon {\cal D} \Lambda_\alpha \quad,
\end{eqnarray}
where $\epsilon$ is a Hermitian Grassmann variable and ${\cal D} =
\partial_\theta + i \theta \partial_t$. As $\{D , {\cal D}\} = 0$, we
also have
\begin{eqnarray}
  \delta (D \Phi^i) & = & -\epsilon {\cal D} (D \Phi^i) \quad, \\
  \delta (D \Lambda^\alpha) & = & -\epsilon {\cal D} (D \Lambda^\alpha) \quad. 
\end{eqnarray}
Using the procedure of Section \ref{constructsect}, we can construct
the fundamental supercharge $Q_0$ from these transformations. This
yields
\begin{equation}
  Q_0 = \pi_i \psi^i - \frac{i}{3} c_{ijk} \psi^i \psi^j \psi^k \quad,
\end{equation}
without any ambiguity arising. The supercharge is generated by
transformations of the dynamical variables $x$, $\psi$ and $\lambda$
so we can check the form of $Q_0$ by calculating the transformations
of $\Phi^i$ and $\lambda^\alpha$ which it generates,
\begin{eqnarray}
  i \epsilon \{Q_0 , \Phi^i\} = - \epsilon {\cal D} \Phi^i & = &
    \delta \Phi^i \label{deltaphi} \quad, \\
  i \epsilon \{Q_0 , \lambda^\alpha\} = - \epsilon F^\alpha & = &
    \delta \lambda^\alpha \label{deltalam} \quad.
\end{eqnarray}
Further, we easily perform a canonical calculation of the Hamiltonian
\begin{equation}
  \{Q_0,Q_0\} = - 2 i H \quad,
\end{equation}
and we find that
\begin{equation}
  H = \frac{1}{2} g_{ij} \dot{x}^i \dot{x}^j - \frac{1}{4} F_{ij
    \alpha \beta} \psi^i \psi^j \lambda^\alpha \lambda^\beta \quad.
    \label{hamiltonian}
\end{equation}
This reproduces, as expected, the canonical result
\begin{equation}
  H = \sum \dot{X} \frac{\partial L}{\partial \dot{X}} - L \quad.
\end{equation}

\subsection{Hidden supersymmetries}
\label{extrasusysect}

To construct further supercharges, we must consider other superfield
transformations which leave the action invariant. Following Coles and
Papadopoulos \cite{ColesandP}, the most general such transformations are
\begin{eqnarray}
  \delta \Phi^i & = & \epsilon {I^i}_j D \Phi^j + i \epsilon
    {e^i}_\alpha \Lambda^\alpha \quad, \label{deltaphi2} \\
  \delta \Lambda^\alpha & = &  \epsilon {I^\alpha}_\beta \nabla
    \Lambda^\beta - {{A_i}^\alpha}_\beta \delta \Phi^i \Lambda^\beta
    - \epsilon {e_i}^\alpha \dot{\Phi}^i \nonumber \\
    & & \mbox{} + i \epsilon {E^\alpha}_{ij} D \Phi^i D \Phi^j + i
    \epsilon {M^\alpha}_{\beta \gamma} \Lambda^\beta \Lambda^\gamma +
    i \epsilon {G^\alpha}_{\beta i} \Lambda^\beta D \Phi^i \quad.
  \label{deltalambda2}
\end{eqnarray}
Here, ${I^i}_j$ and ${I^\alpha}_\beta$ are endomorphisms of the sigma
model manifold $\M$ and the fibre $G$ respectively; ${e^i}_\alpha$ and
${e_i}^\alpha$ are bundle maps between the manifold and the fibre. In
the case where $\dim \M = \dim G$ and $h_{\alpha \beta}$ is the flat
metric on $G$, the $e$ can be interpreted as vielbeine. It should be
noted that terms in ${I^\alpha}_\beta$ do not appear in $\delta
\lambda_\alpha$.  Thus the ${I^\alpha}_\beta$ do not affect the
dynamical variables and consequently do not appear in the
supercharges. This explains the absence of terms involving
${I^\alpha}_\beta$ from the conditions below. As described in Section
\ref{constructsect}, we can obtain a set of conditions, such as those
presented in \cite{ColesandP}, {\it sufficient} for the invariance of
the action under (\ref{deltaphi2}) and (\ref{deltalambda2}), these being
determined {\it only up to partial integrations in
superspace}. Demanding that the supercharges we construct do generate
the original supersymmetry transformations requires the set of
conditions on the fields appearing in (\ref{deltaphi2}) and
(\ref{deltalambda2}) to be in {\it exactly} the following form:

\vspace{.1in} \noindent Conditions associated with $I$
\begin{eqnarray}
  I_{(ij)} & = & 0 \quad, \label{1stcondition} \\
  \nabla_i I_{jk} + \nabla_{[k} I_{j]i} + \frac{3}{2} c_{mjk} {I^m}_i
    & = & 0 \quad, \label{2ndIcond} \\
  {I^m}_{[i} F_{j]m \alpha \beta} & = & 0 \quad, \\
  G_{\alpha \beta i} & = & 0 \quad.
\end{eqnarray}
\vspace{.1in} \noindent Conditions associated with $e$
\begin{eqnarray}
  h_{\alpha \beta} {e_i}^\beta - g_{ij} {e^j}_\alpha & = & 0 \quad, \\
  M_{\alpha \beta \gamma} - M_{[\alpha \beta \gamma]} & = & 0 \quad, \\
  {e^k}_\alpha c_{kij} + \nabla_i e_{j\alpha} + 2 E_{\alpha ij} & =
    & 0 \quad, \label{3rdecond} \\
  \nabla_{[i}\left( c_{jk]m} {e^m}_\alpha \right) + \frac{1}{2}
  {e_{[i}}^\beta F_{jk]\alpha \beta} & = & 0 \quad, \label{4thecond} \\
  \frac{2}{3} \nabla_i M_{\alpha \beta \gamma} + F_{ij[\alpha \beta}
    {e^j}_{\gamma ]} & = & 0 \quad. \label{lastcondition}
\end{eqnarray}
In particular, we note the results, from (\ref{2ndIcond}) and
(\ref{3rdecond}) respectively, that
\begin{eqnarray}
  c_{n[ij} {I^n}_{k]} & = & 0 \quad, \label{simpcond1} \\
  \nabla_{(i} e_{j) \alpha} & = & 0 \quad. \label{simpcond2}
\end{eqnarray}
To construct the supercharges, we proceed as above. However, the
transformations decouple into the parts generated by the endomorphisms
$I$ and the parts generated by the bundle maps $e$ (and the three-form
$M$). These can therefore be treated independently in the construction
of supercharges. We will call these supersymmetries type I and type II
respectively. We will use the notation $Q_1(I)$ for a type I
supersymmetry generated by $I$ and $Q_2(e,M)$ for a type II
supersymmetry generated by $e$ and $M$. Thus, to construct $Q_1(I)$,
we use
\begin{eqnarray}
  \delta \Phi^i & = &  \epsilon {I^i}_j D \Phi^j \quad, \nonumber \\
  \delta \Lambda^\alpha & = &  \epsilon {I^\alpha}_\beta \nabla
    \Lambda^\beta - {{A_i}^\alpha}_\beta \delta \Phi^i \Lambda^\beta
    \nonumber \quad,
\end{eqnarray}
which give the type I supercharge
\begin{equation}
  Q_1(I) = \pi_i {I^i}_j \psi^j - \frac{i}{3} \left( \nabla_i I_{jk}
    \right) \psi^i \psi^j \psi^k \quad.
  \label{Q1def}
\end{equation}
To construct $Q_2(e,M)$, we use
\begin{eqnarray}
  \delta \Phi^i & = & i \epsilon {e^i}_\alpha \Lambda^\alpha \nonumber
    \quad, \\
  \delta \Lambda^\alpha & = & - {{A_i}^\alpha}_\beta \delta \Phi^i
    \Lambda^\beta - \epsilon {e_i}^\alpha \dot{\Phi}^i + i \epsilon
    {E^\alpha}_{ij} D \Phi^i D \Phi^j + i \epsilon {M^\alpha}_{\beta
    \gamma} \Lambda^\beta \Lambda^\gamma \nonumber \quad,
\end{eqnarray}
which give the type II supercharge
\begin{equation}
  Q_2(e,M) = \pi_i {e^i}_\alpha \lambda^\alpha + i E_{\alpha ij}
    \lambda^\alpha \psi^i \psi^j - \frac{i}{3} M_{\alpha \beta
    \gamma} \lambda^\alpha \lambda^\beta \lambda^\gamma \quad.
  \label{Q2def}
\end{equation}
At this point, we can verify the correctness of our result by showing
that the canonical brackets of $Q_1$ and $Q_2$ with the dynamical
variables do indeed generate the required transformations of these
variables, as in (\ref{deltaphi}) and (\ref{deltalam}). Non-trivial
calculations allow the following results to be verified 
\begin{eqnarray}
  \{Q_0 , Q_1(I)\} & = & 0 \quad, \\
  \{Q_0 , Q_2(e,M)\} & = & 0 \quad. 
\end{eqnarray}
Of course, it should be emphasised that, in presenting the results
(\ref{1stcondition}) to (\ref{lastcondition}), we have already fixed the
detail in them so that the consistency arguments just described work
correctly. Results (\ref{Q1def}) and (\ref{Q2def}) are new here. If one
specialises the results of \cite{ColesandP} to our (somewhat less
general) context, we see that (\ref{2ndIcond}) and (\ref{4thecond})
contain important refinements of these results. These refinements are
critical to the problem of determining the supercharges associated to
the extra supersymmetries which we describe.

\section{A simplification}
\label{simpsect}

If we are prepared to break manifest covariance in the expressions for
the supercharges then it is possible to simplify some subsequent
calculations significantly. To this end we define $\tilde{\pi}$, by
\begin{eqnarray}
  \tilde{\pi}_i & = & \pi_i - \mbox{\{$c$ terms \}} \quad, \\
  \tilde{\pi}_i & = & p_i - \frac{i}{2} g_{ij,k} \psi^j \psi^k -
    \frac{i}{2} A_{i \alpha \beta} \lambda^\alpha \lambda^\beta \quad,
\end{eqnarray}
so that, upon use of (\ref{3rdecond}) and (\ref{simpcond1}), we obtain
\begin{eqnarray}
  Q_0 & = & \tilde{\pi}_i \psi^i + \frac{i}{6} c_{ijk} \psi^i \psi^j
    \psi^k \quad, \\
  Q_1 & = & \tilde{\pi}_i {I^i}_j \psi^j - \frac{i}{3} \left( \nabla_i
    I_{jk} \right) \psi^i \psi^j \psi^k \quad, \label{simpQ1} \\
  Q_2 & = & \tilde{\pi}_i {e^i}_\alpha \lambda^\alpha - \frac{i}{2} \left(
    \nabla_i e_{j \alpha} \right) \psi^i \psi^j \lambda^\alpha -
    \frac{i}{3} M_{\alpha \beta \gamma} \lambda^\alpha \lambda^\beta
    \lambda^\gamma \quad. \label{simpQ2} 
\end{eqnarray}
In terms of $x$, $\tilde{\pi}$, $\psi$ and $\lambda$, the general
Dirac bracket simplifies to
\begin{eqnarray}
  \{A,B\} & = & \frac{\partial A}{\partial x^i} \frac{\partial
    B}{\partial \tilde{\pi}_i} -\frac{\partial A}{\partial \tilde{\pi}_i}
    \frac{\partial B}{\partial x^i} \nonumber \\
  & + & \frac{\partial A}{\partial \tilde{\pi}_i} \frac{\partial B}{\partial
    \tilde{\pi}_j} \left( \frac{i}{2} R_{ijpq} \psi^p \psi^q +
    \frac{i}{2} F_{ij \alpha \beta} \lambda_\alpha \lambda_\beta
    \right) \nonumber \\
  & - & \left({(-)}^b \frac{\partial A}{\partial \tilde{\pi}_i}
    \frac{\partial B}{\partial \psi^j} - {(-)}^{a+b} \frac{\partial
    A}{\partial \psi^j} \frac{\partial B}{\partial \tilde{\pi}_i} \right)
    {\Gamma^j}_{ik} \psi^k \nonumber \\
  & - & \left( {(-)}^b \frac{\partial A}{\partial \tilde{\pi}_i} \frac{\partial
    B}{\partial \lambda_\alpha} - {(-)}^{a+b} \frac{\partial
    A}{\partial \lambda_\alpha} \frac{\partial B}{\partial \tilde{\pi}_i}
    \right) A_{i \alpha \beta} \lambda_\beta \nonumber \\
  & + & i {(-)}^a \frac{\partial A}{\partial \psi^i} \frac{\partial
    B}{\partial \psi^j} g^{ij} + i {(-)}^a \frac{\partial A}{\partial
    \lambda_\alpha} \frac{\partial B}{\partial \lambda_\beta}
    h^{\alpha \beta} \quad.
\end{eqnarray}
It is helpful and appropriate to employ (\ref{simpQ1}) and
(\ref{simpQ2}) in the discussion of the supercharge algebra via the
classical Poisson-Dirac bracket. We then have the conditions:

\vspace{.1in} \noindent Commutation of two type I supercharges;
$\{Q_1(I)\, ,\,Q_1'(J)\}=0$: 
\begin{eqnarray}
  I_{m(i} {J_{j)}}^m &=& 0 \quad, \\
  {I^m}_{[i} {J^n}_{j]} F_{mn\alpha \beta} &=& 0 \quad, \\
  {{\cal N}(I,J)^i}_{jk} \equiv \left(I_{m[j} \nabla^m {J^i}_{k]} -{I^i}_m
    \nabla_{[j} {J^m}_{k]}\right) + (I \leftrightarrow J) &=& 0 \quad, \\
  R_{mn[ij} {I^m}_k {J^n}_{l]} &=& 0 \quad, \\
  J_{m[i} \nabla^m \gamma^{(I)}_{jkl]}+\frac{3}{2} \gamma^{(I)}_{m[ij}
    \gamma^{(J)m}_{kl]} + (I \leftrightarrow J, \gamma^{(I)} \leftrightarrow
    \gamma^{(J)}) &=& 0 \quad,
\end{eqnarray}
where $\gamma^{(I)}_{ijk} = \nabla_{[i}I_{jk]}$. The $(1,2)$-tensor
${{\cal N}^i}_{jk}$ is the Nijenhuis concomitant of the two
endomorphisms $I$ and $J$.

\vspace{.1in} \noindent Commutation of two type II supercharges;
$\{Q_2(e,M)\, ,\,Q_2'(f,N)\}=0$: 
\begin{eqnarray}
  {e^{(i}}_\alpha f^{j)\alpha} &=& 0 \quad, \\
  M_{\mu[\alpha \beta} {N_{\gamma\sigma]}}^\mu - \frac{1}{2}
    F_{ij[ \alpha \beta} {e^i}_\gamma {e^j}_{\sigma]} &=& 0 \quad, \\
  f_{n[\alpha} \nabla^n {e^i}_{\beta]} +f^{i\mu} M_{\mu \alpha \beta}
    + (e \leftrightarrow f, M \leftrightarrow N) &=& 0 \quad, \\
  {e_{[i}}^\alpha \nabla_j f_{k]\alpha} + {f_{[i}}^\alpha \nabla_j
    e_{k]\alpha} &=& 0 \quad, \\
  {e_{[i}}^\alpha {f_j}^\beta F_{kl]\alpha \beta} &=& 0 \quad, \\
  {f_{[i}}^\gamma F_{j]k \alpha[\beta} {e^k}_{\gamma]} + (e \leftrightarrow f)
  &=& 0 \quad. 
\end{eqnarray}
\vspace{.1in} \noindent Commutation of type I and type II supercharges;
$\{Q_1(I)\, ,\,Q_2(e,M)\}=0$:
\begin{eqnarray}
  {e^k}_\alpha \nabla_k I_{ij} + 2 I_{k[i} \nabla^k e_{j] \alpha} & =
    & 0 \quad, \label{Q1Q2cond1} \\
  {I^n}_{[i} F_{jk] \alpha \beta} e_{n \beta} & = & 0 \quad. 
  \label{Q1Q2cond2}
\end{eqnarray}
We look for the maximal commuting set $\{ Q_{1i} \, ,\, Q_{2j} \}$ as
the maximal extension of the supersymmetry algebra.

\section{An Example}
\label{examplesect}

We describe here the treatment of an explicit example in order to
illustrate the subtleties associated with the construction of the
supercharges that are the central focus of this paper. Consider the
special case of (\ref{superL}) 
\begin{equation}
  S_0 =  \int \,d \theta \,dt \frac{1}{2} i g_{ij}(\Phi) \dot{\Phi^i} D
    \Phi^j \quad,
\end{equation}
and (\ref{deltaphi2})
\begin{equation}
  \delta \Phi^i = \epsilon {I^i}_j (\Phi) D \Phi^j \quad.
  \label{exdeltaphi}
\end{equation}
We compute $\delta S_0$ directly, finding five terms. We first treat the
two terms which do not involve any derivatives of $g_{ij}$ or
$I_{ij}$. Using an integration by parts to derive the second line, we
find
\begin{eqnarray}
  \delta S_0 &=& \frac{1}{2} i \int \,d \theta \,dt \epsilon \left[
    \ldots I_{ij}(\Phi) D \dot{\Phi}^j D\Phi^i - I_{ij}(\Phi) \dot{\Phi}^i
    D^2 \Phi^j + \ldots \right] \label{exdeltaS1} \\
  &=& \frac{1}{2} i \int \,d \theta \,dt \epsilon \left[ \ldots D \left(
    I_{ij}(\Phi) D \Phi^i \dot{\Phi}^j \right) -I_{ij,k}(\Phi) D\Phi^k
    D\Phi^i \dot\Phi^j \right] \quad. \label{exdeltaS2}
\end{eqnarray}
To reach this point, we have eliminated terms of the form $I_{ij}
\dot{\Phi}^i \dot{\Phi}^j$ by imposing the condition
\begin{equation}
  I_{ij} + I_{ji} = 0 \quad,
 \label{Icond1}
\end{equation}
where $I_{ij} = g_{ik} {I^k}_j$. The divergence term in
(\ref{exdeltaS2}) does not contribute to $\delta S_0$ but does
contribute to $\delta L_0$ and hence, via Noether's theorem, to the
supercharge $\tilde{Q}$ associated with (\ref{exdeltaphi}). We next
collect the remaining terms so as to absorb derivatives of $g_{ij}$ into
Christoffel symbols, obtaining
\begin{equation}
  \delta S_0 = -\frac{1}{2} \epsilon \int \,d \theta \,dt \left( I_{jk,i} - 2
    {\Gamma^p}_{ij} I_{pk} \right) D\Phi^j D \Phi^k D^2 \Phi^i \quad.
  \label{exdeltaS}
\end{equation}
Demanding that the part of the bracket in (\ref{exdeltaS}) antisymmetric
in $j$ and $k$ vanishes is sufficient to ensure that $\delta S_0$
vanishes. However, a direct calculation of the supercharge $\tilde{Q}$
using Noether's theorem leads to a form of $\tilde{Q}$ that fails to
reproduce the original transformation (\ref{exdeltaphi})
canonically. The nature of the failure prompts us to split the first
term of (\ref{exdeltaS}) using the identity
\begin{equation}
  I_{jk,i} D \Phi^j D \Phi^k D^2 \Phi^i = \frac{1}{3} D \left[
    I_{jk,i} D \Phi^j D \Phi^k D \Phi^i \right] + \frac{2}{3} \left(
    I_{jk,i} + I_{ji,k} \right) D \Phi^j D \Phi^k D^2 \Phi^i \quad,
\end{equation}
bringing in a total derivative term of the type  that is needed
to improve (and, it turns out, to correct) the Noether expression for 
$\tilde{Q}$. Again, the total derivative term (which is merely the
result of an integration by parts) does  not contribute to $\delta S_0$
but does contribute to $\tilde{Q}$. It now follows that $\delta S_0$,
including the connection term from (\ref{exdeltaS}), is given by  
\begin{equation}
  -\frac{1}{3} \epsilon \int \,d \theta \,dt \left[ \nabla_i I_{jk} +
    \frac{1}{2} \nabla_k I_{ji} - \frac{1}{2} \nabla_j I_{ki} \right] D 
    \Phi^k D
    \Phi^j D^2 \Phi^i \quad,
\end{equation}
where $\nabla$ is the metric-covariant derivative. Thus $\delta S_0 =0$ can be
realised by imposing the condition
\begin{equation}
  \nabla_i I_{jk} + \nabla_{[k} I_{j]i} = 0 \quad.
  \label{exIcond}
\end{equation}
This result, contained in the condition (\ref{2ndIcond}) arising in the
general case above, represents a crucial modification of the
corresponding equation of \cite{ColesandP}. 

The supercharge $\tilde{Q}$ as defined in (\ref{schargedef}) is
\begin{equation}
  i \epsilon \tilde{Q} = \sum_X \delta X \frac{\partial L}{\partial
    \dot{X}} - J \quad, 
  \label{exQdef}
\end{equation}
where the sum is over all dynamical variables and $J$ is as calculated
above, that is
\begin{eqnarray}
  J &=& \epsilon \int \,d \theta \,dt \left[-\frac{i}{2} D\left(I_{ij}
    (\Phi) D \Phi^i \dot{\Phi}^j \right) -\frac{1}{6} D \left( \nabla_i I_{jk}
    (\Phi) D\Phi^i D\Phi^j D\Phi^k \right) \right] \quad, \\
  &=& \frac{i \epsilon}{2} I_{ij}(x) \dot{x}^i \psi^j +\frac{ \epsilon}{6}
    \nabla_i I_{jk}(x) \psi^i \psi^j \psi^k \quad.
\end{eqnarray}
Explicit calculation of the expression (\ref{exQdef}) yields the
following expression for the supercharge $\tilde{Q}$, 
\begin{equation}
  \tilde{Q} = I_{ij}(x) \dot{x}^i \psi^j -\frac{i}{3} \nabla_i I_{jk}(x)
    \psi^i \psi^j \psi^k \quad. 
\end{equation}   
in agreement with that presented above in (\ref{Q1def}). Finally, use of
the canonical formalism of the theory (see Section \ref{canonsect})
allows us to verify the central result that, with precisely the
condition (\ref{exIcond}) on $I_{ij}$, the supercharge $\tilde{Q}$ does
indeed generate the original supersymmetry transformation
(\ref{exdeltaphi}), that is
\begin{equation}
  -i\epsilon \{Q,\Phi^i \} = \epsilon {I^i}_j D \Phi^j = \delta \Phi^i
    \quad. 
  \label{exQtrans}
\end{equation}
Of course, one does not know that the job of determining the exact 
conditions that must be imposed upon ${I^i}_j$ is indeed complete until
a Noether charge has been computed and seen to satisfy
(\ref{exQtrans}). The results of (\ref{1stcondition}) to
(\ref{lastcondition}) were in fact obtained by generalising the
procedure followed in this Section, being so arranged as yield Noether
charges which generate (\ref{deltaphi2}) and (\ref{deltalambda2})
canonically.

\section{Conclusion}

We have considered the $N=1$ supersymmetric nonlinear
sigma-model and described the conditions under which extra
supersymmetries of the most general type can exist. We have derived the
conditions for invariance of the action, which were defined up to
partial integrations in superspace and shown that there is a unique
form of these which is required for the construction, by Noether's
theorem, of the supercharges. The precise form of these is determined by
imposing the necessary requirement that the supercharges generate the original
supersymmetry transformations. We explicitly constructed supercharges
for this model and investigated their algebra via the canonical
Poisson-Dirac bracket of the theory.

\end{document}